\documentclass[conference]{IEEEtran}

\usepackage{setspace}
\usepackage{multirow}
\usepackage[export]{adjustbox}

\IEEEoverridecommandlockouts
\usepackage{cite}
\usepackage{amsmath,amssymb,amsfonts}
\usepackage{algorithmic}
\usepackage{graphicx}
\usepackage{textcomp}
\usepackage{xcolor}

\def\BibTeX{{\rm B\kern-.05em{\sc i\kern-.025em b}\kern-.08em
    T\kern-.1667em\lower.7ex\hbox{E}\kern-.125emX}}

\begin{document}

\title{Gathering Cyber Threat Intelligence from Twitter Using Novelty Classification
}

\author{\IEEEauthorblockN{Ba-Dung Le}
\IEEEauthorblockA{\textit{School of Computer Science} \\
\textit{University of Adelaide}\\
Adelaide, Australia \\
badung.le@adelaide.edu.au}
\and
\IEEEauthorblockN{Guanhua Wang}
\IEEEauthorblockA{\textit{School of Computer Science} \\
	\textit{University of Adelaide}\\
	Adelaide, Australia \\
	guanhua.wang@adelaide.edu.au}
\and
\IEEEauthorblockN{Mehwish Nasim}
\IEEEauthorblockA{\textit{School of Mathematical Sciences} \\
\textit{University of Adelaide}\\
Adelaide, Australia \\
mehwish.nasim@adelaide.edu.au}
\and
\IEEEauthorblockN{M. Ali Babar}
\IEEEauthorblockA{\textit{School of Computer Science} \\
\textit{University of Adelaide}\\
Adelaide, Australia \\
ali.babar@adelaide.edu.au}

}

\maketitle

\begin{abstract}
Preventing organizations from Cyber exploits needs timely intelligence about Cyber vulnerabilities and attacks, referred to as threats.
Cyber threat intelligence can be extracted from various sources including social media platforms where users publish the threat information in real-time.
Gathering Cyber threat intelligence from social media sites is a time-consuming task for security analysts that can delay timely response to emerging Cyber threats.
We propose a framework for automatically gathering Cyber threat intelligence from Twitter by using a novelty detection model.
Our model learns the features of Cyber threat intelligence from the threat descriptions published in public repositories such as  Common Vulnerabilities and Exposures (CVE) and classifies a new unseen tweet as either normal or anomalous to Cyber threat intelligence.
We evaluate our framework using a purpose-built  data set of tweets from 50 influential Cyber security-related accounts over twelve months (in 2018). 
Our classifier achieves the F1-score of 0.643 for classifying Cyber threat tweets and outperforms several baselines including binary classification models. 
Analysis of the classification results suggests that Cyber threat-relevant tweets on Twitter do not often include the CVE identifier of the related threats. Hence, it would be valuable to collect these tweets and associate them with the related CVE identifier for Cyber security applications.
\end{abstract}


\section{Introduction}
Recently, there has been an increasing reliance on the Internet for business, government, and social interactions as a result of a trend of hyper-connectivity and hyper-mobility.
While the Internet has become an indispensable infrastructure for businesses, governments, and societies, there is also an increased risk of Cyber attacks with different motivations and intentions. For examples, a U.S. government report \cite{Ransomware} shows that there was an average of more than 4000 ransomware attacks per day in 2016 - a four fold increase compared to 2015. 
According to Cybersecurity Ventures \cite{cybersecurityventures}, Cyber crime will continue to rise with a combined cost to businesses globally more than \$6 trillion annually by 2021. Therefore, Cyber security has become a critically important area of research and practice over the last few years. 

Preventing organizations from Cyber exploits needs timely intelligence about Cyber vulnerabilities and attacks, referred to as threats. Threat intelligence is defined as ``evidence-based knowledge, including context, mechanisms, indicators, implications and actionable advice, about an existing or emerging menace or hazard to assets that can be used to inform decisions regarding the subject's response to that menace or hazard" \cite{gartner2013}. Threat intelligence in Cyber security domain, or Cyber threat intelligence, provides timely and relevant information, such as signatures of the attacks, that can help reduce the uncertainty in identifying potential security vulnerabilities and attacks. 

\begin{figure*}[h!]		
	\begin{center}		
		\includegraphics[scale=0.95,frame]{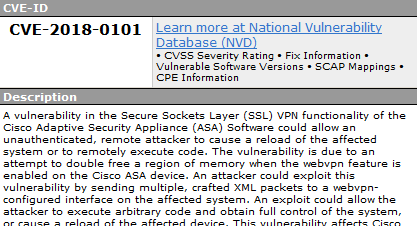}	
	\end{center}
	\caption{An example of the entries in the CVE database}
	\label{fig:CVE-example}
\end{figure*}

Cyber threat intelligence can generally be extracted from overt or formal sources, which officially release threat information in structured data format. Structured threat intelligence adhere to a well-defined data model, with common format and structure, such as an XML schema. Structured Cyber threat intelligence, therefore, can be easily parsed by security tools to analyze and respond to security threats accordingly. 
Examples of formal sources of Cyber threat intelligence include the Common Vulnerabilities and Exposures (CVE) database \cite{mitre} and the National Vulnerability Database (NVD) \cite{nvd}. 
Fig.~\ref{fig:CVE-example} shows an example of the entries  in the CVE database relating to a threat. Each CVE entry has an identifier (ID) that includes the prefix `CVE', the year that the CVE entry was created or published and a sequence number of four or more digits. 
A CVE entry also has a brief description of the threat that generally includes the information about the affected product, versions and vendor, the threat type and the impact, method and inputs of an attack. However, some of these details may not be included in a CVE description if the information is not available at the publishing time. 

Cyber threat intelligence is also available on covert or informal sources, such as public blogs, dark webs, forums and social media platforms.
Informal sources allow any person or entity on the Internet to publish, in real-time, the threat information in natural language, or unstructured data format. The unstructured and publicly available threat intelligence are also called Open Source Intelligence (OSINT) \cite{steele1996open}. 
Cyber security related OSINT are early warning sources for Cyber security events such as security vulnerability exploits \cite{sabottke2015vulnerability}.
For examples, in June 2017, the global ransomware outbreak of `Petya/NotPetya' was discussed widely via Twitter before being reported by mainstream media \cite{sapienza2018discover}.
To prioritize response to Cyber threats, Cyber security analysts must quickly determine the emerging threats that are currently discussed on public sources. However, gathering Cyber OSINT is a time-consuming task as natural language is ambiguous and difficult for security tools to parse. Any delay in taking suitable actions against a security vulnerability, threat, or attack can lead to more loss.

\begin{figure*}[h!]		
	\begin{center}
		\includegraphics[scale=0.85, frame]{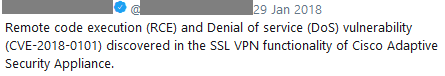}
		\includegraphics[scale=0.86,frame]{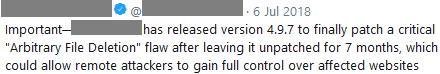}
		\includegraphics[scale=0.85,frame]{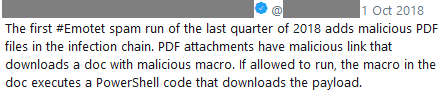}	
	\end{center}
	\caption{Examples of the tweets about Cyber threats}
	\label{tbl: threat-tweet-CVE}
\end{figure*}

The work reported in this paper has focused on collecting and analyzing data from Twitter, which allows its users to post 280 character long messages, called tweets. 
Twitter is a main source for Cyber OSINT as many Cyber security experts are using this open platform to disseminate information about Cyber threats \cite{queiroz2017predicting}.
Fig.~\ref{tbl: threat-tweet-CVE} shows a few examples of Cyber threat-relevant tweets on Twitter. 
The first tweet summarizes the CVE entry with the identifier `CVE-2018-0101'. 
The second and third tweets discuss two different threats but do not include any CVE identifier. However, using our  knowledge about Cyber security, we can associate these two tweets with the CVE identifiers `CVE-2018-20714' \footnote{https://thehackernews.com/2018/06/wordpress-hacking.html}  and `CVE-2017-11882'  \footnote{https://www.trendmicro.com/vinfo/au/security/news/vulnerabilities-and-exploits/17-year-old-ms-office-flaw-cve-2017-11882-actively-exploited-in-the-wild} respectively.
Collecting these tweets with the associated CVE identifiers is useful for Cyber threat-related applications such as exploit prediction \cite{sabottke2015vulnerability} and Indicators of Compromise (IoCs) generation \cite{dionisio2019cyberthreat}. 

We have developed a framework for automatically gathering Cyber threat intelligence from Twitter. Our framework utilizes a novelty detection model to classify the tweets as relevant or irrelevant to Cyber threat intelligence. The novelty classifier learns the features of Cyber threat intelligence from the threat descriptions in the CVE database and classifies a new unseen tweet as normal or abnormal in relation to Cyber threat intelligence.
The normal tweets are considered as Cyber threat-relevant while the abnormal tweets are considered as Cyber threat-irrelevant.
We evaluate our framework on a purpose-built data set created from the tweets collected over a period of twelve months in 2018 from 50 influential Cyber security related accounts. 
During the evaluation, our framework achieved the highest performance of 0.643 measured by the F1-score metric for classifying Cyber threat tweets. To our knowledge, our approach outperformed several baselines including binary classification models. 
We have analyzed the correctly classified Cyber threat tweets and discovered that 81 of them do not contain a CVE identifier.
We have also found that 34 out of the 81 tweets can be associated with a CVE identifier included in the top 10 most similar CVE descriptions of each tweet.

The highlights of this work are:

\begin{itemize}
	\item An automated framework for detecting Cyber threat tweets on Twitter using novelty classification
	\item An evaluation of our framework on a challenging data set created from the tweets collected over a period of twelve months from 50 influential Cyber security related accounts
	\item A detailed description of an analysis and the results of the relationship between the correctly classified Cyber threat tweets and threat descriptions in the CVE database
\end{itemize}

The rest of the paper is organized as follows.
In section 2, we summarize the existing work related to automatically gathering Cyber OSINT from Twiter. In section 3, we present our framework for the automated collection task. We evaluate our framework and discuss our findings in section 4. In section 5, we presents our conclusions from the results of our work and suggests directions for future work. 

\section{Related Work}
In the last few years, research on using Cyber threat-relevant information available on Twitter for security purposes has gained significant attention. 
To automatically collect Cyber threat intelligence from Twitter, several methods have been used \cite{sabottke2015vulnerability,le2017sonar,trabelsi2015mining,mittal2016cybertwitter,sapienza2018discover,alves2019processing,dionisio2019cyberthreat}.

The most traditional method for collecting Cyber threat-relevant tweets is searching for the tweets containing the CVE identifier \cite{sabottke2015vulnerability}. Sabottke at. al.  \cite{sabottke2015vulnerability} use this collection method for predicting Cyber exploits in the real world. Their exploit detector uses the collected Cyber threat tweets to improve the precision of the prediction model and to generate early exploit warnings. 
However, because the tweets that do not contain the CVE identifier are ignored, their exploit detector might not appropriately take into account the potential exploits relating to the ignored Cyber threat tweets.

Le Sceller at. al. \cite{le2017sonar} collect Cyber threat information, referred to as Cyber security events, on Twitter based on a set of related keywords. Cyber threat irrelevant information, that might have been collected, are discarded using backlist keywords. Over time, new related keywords are added into the set of the initially related keywords using a self-learned mechanism.
Sapienza et al. \cite{sapienza2018discover} identify Cyber threat tweets as the tweets containing a number of terms in a set of Cyber security related terms. Trabelsi et. al. \cite{trabelsi2015mining} collect Cyber threat tweets based on both the CVE identifier and a set of Cyber security related keywords. 
Mittal et al. \cite{mittal2016cybertwitter} combine the keywords based collection method and Name Entity Recognition (NER) to collect Cyber threat information.
The drawback of the keywords based collection method for Cyber threat information is that this method requires expert knowledge about Cyber threats to choose the relevant keywords. The keywords based collection method, therefore, can easily ignore Cyber threat-related information and collects Cyber threat irrelevant information if the keywords are not carefully selected \cite{le2017sonar}. 

Alves at. al. \cite{alves2019processing} focus on designing a completed online monitoring system for Cyber threat tweets on Twitter. Their monitoring system includes a Cyber threat tweet classification module that uses supervised machine learning approach to classify Cyber threat tweets. This module  transforms tweets to vector representations and classifies the tweets as Cyber threat relevant or irrelevant using binary classification models, particularly Support Vector Machines (SVM) and Multi-Layer Perceptron (MLP) neural networks.
Dionísio et. al. \cite{dionisio2019cyberthreat} use word embeddings such as GloVE \cite{pennington2014glove} and Word2Vec \cite{le2014distributed} for feature extraction and use the binary classification model Convolutional Neural Network (CNN) \cite{kim2014convolutional} for classifying Cyber threat tweets.
The collection method for Cyber threat tweets based on binary classification requires the classifiers  to be trained with both positive and negative samples, or Cyber threat-relevant and Cyber threat-irrelevant tweets. 
This potentially introduces the problem of sampling bias
which occurs when the positive or negative samples are not the representative of Cyber threat-relevant or Cyber threat-irrelevant tweets respectively.

\section{Gathering Cyber Threat Tweets Using Novelty Classification}

\begin{figure*}[h!]
	\begin{center}
		\includegraphics[]{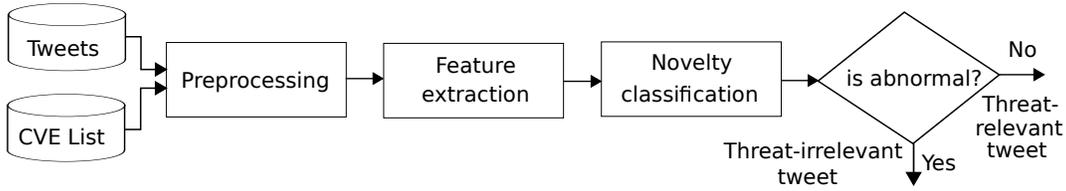}	
		\caption{Architecture of our framework for classifying Cyber threat tweets}
		\label{fig_1}
	\end{center}
\end{figure*}

As previously reported, our work focuses only on the collection method of Cyber threat tweets instead of a complete system with functional requirements such as scalability, real-time processing and security alert generation as in some previous work \cite{mittal2016cybertwitter,alves2019processing,dionisio2019cyberthreat}. 
The key idea of our method is that we formulate the task of detecting Cyber threat tweets as a novelty classification task \cite{scholkopf2000support}. 
A novelty classifier needs to be trained only with  positive samples without using negative samples. 
After being trained, the novelty classifier subsequently applies its knowledge to decide whether a new unseen tweet is normal or abnormal to the class of the positive samples. By using novelty classification, we avoid the issue of sampling bias toward the negative training data set.

Fig.~\ref{fig_1} shows the architecture of our framework for classifying Cyber threat tweets. 
Our framework consists of three phases including pre-processing, feature extraction and novelty classification. 
The input of our framework includes the tweets collected from Twitter and the threat descriptions from the CVE database \cite{mitre}. The CVE descriptions are used as the positive samples for training our novelty classifier. 
The output of our framework consists of the tweets that are classified as normal, or Cyber threat-relevant, and the tweets that are classified as abnormal, or Cyber threat-irrelevant.

\subsection{Preprocessing}
The preprocessing phase is to eliminate the terms in the input documents that are unnecessary for identifying Cyber threat information. 
This phase converts the input documents into lowercase with punctuation, numbers, hyperlinks, mentions and hashtags stripped out. 
Stopwords in the input documents are also removed using the default stopword list in the Natural Language Toolkit (NLTK) package \footnote{The NLTK package can be downloaded at https://www.nltk.org/}. 
We do not apply stemming and lemmatizing onto the input documents as it may change the meaning of them.


\subsection{Feature extraction}
The feature extraction phase is to transform the pre-processed documents into numerical vector representations for classification.
To represent each document as a vector, we use the Term Frequency-Invert Document Frequency (TF-IDF) method \cite{joachims1996probabilistic,salton1988term} which assigns weights to the document terms as follows.
Let d is a document in a corpus and t is a term in the document. The weight of term t in document d is defined as
$$TF-IDF(t,d) = f(t,d)*log(N/n_t),$$
where 
$f(t,d)$ is the number of the occurrences of term t in document d,
N is the total number of documents in the corpus
and $n_t$ is the number of the documents containing term t. 

It is noted that our training corpus consists of only positive samples. Therefore, the total number of documents in our training corpus is the total number of the positive samples.

\subsection{Novelty classification}
After transforming the collected tweets and the CVE descriptions into numerical vectors, we use a novelty classifier to classify each of the input tweets as normal or abnormal to the class of Cyber threat intelligence. 
To choose a suitable classification model, we explore two different novelty classifiers including Centroid \cite{grubbs1969procedures,guan2009class} and One-class Support Vector Machine \cite{scholkopf2000support,manevitz2001one}. 


The Centroid classifier \cite{grubbs1969procedures,guan2009class} decides whether an input document is normal or abnormal to the positive class based on the distance between the input document and the centroid of the positive class. The centroid C of a class S of documents is defined as 
$$C=\frac{1}{|S|} \sum_{d \in S}v_d,$$
where d is a document in S,
$v_d$ is the vector representation of document d
and $\mid$S$\mid$ is the total number of document in S.

Given a threshold value, an input document is classified abnormal to the positive class if the distance between the document and the centroid is larger than the threshold value. Otherwise, the input document is classifier as normal to the positive class. 
The distance between two vectors $v_i$ and $v_j$ is computed as the cosine similarity between the two vectors, which is defined as
$$cos(v_i, v_j)=\frac{v_i. v_j}{||v_i||*||v_j||}.$$

The One-class Support Vector Machine (One-class SVM) classifier \cite{scholkopf2000support,manevitz2001one} aims at finding a function that returns a positive value for a normal data point of the positive class and a negative value for an abnormal data point.
As finding the function is difficult in the original feature space, the One-class SVM classifier maps the input data points into a high dimensional feature space via a kernel. 
The mapping kernel transforms the abnormal or novel data point closer to the origin than the members. 
The One-class SVM classifier then finds the hyperplane that separates the training class from the origin with maximum margin. 
For an input data point, the function returns a value deciding the side of the hyperplane that the input data point falls on.
We use the implementation of One-class SVM classifier in the scikit-learn Python package \footnote{The scikit-learn Python package can be downloaded at https://scikit-learn.org/}.

\section{Performance evaluation}
\subsection{Experiment setting}
\paragraph{Training and testing data sets}

\begin{table*}[h!]
	\caption{List of the 50 Twitter accounts for collecting Cyber threat tweets}
	\label{tbl:accounts}
	\begin{center}
		\begin{tabular}{|p{15cm}|}			
			\hline
			\begin{small}				
				avast\_antivirus,
				cyber,
				CyberSec\_\_News,
				MalwareTechBlog,
				lennyzeltser,
				securityaffairs,
				CSOonline,
				DarkReading,
				helpnetsecurity,
				USCERT\_gov,
				Peerlyst,
				e\_kaspersky,
				troyhunt,
				jeremiahg,
				schneierblog,
				mikko,
				IBMSecurity,
				k8em0,
				briankrebs,
				OracleSecurity,
				TenableSecurity,
				Cybersec\_EU,
				Hacker\_Combat,
				securityonion,
				AdobeSecurity,
				circl\_lu,
				USCyberMag,
				Secureworks,
				WDSecurity,
				CiscoSecurity,
				CarbonBlack\_Inc,
				MISPProject,
				Binary\_Defense,
				FireEye,
				EmergingThreats,
				InfosecurityMag,
				EHackerNews,
				TheHackersNews,
				TrendMicro,
				SecurityWeek,
				Sophos,
				threatintel,
				NortonOnline,
				McAfee,
				symantec,
				kaspersky,
				RecordedFuture,
				alienvault,
				Unit42\_Intel,
				CyberGovAU	
			\end{small}\\
			\hline
		\end{tabular}
	\end{center}
\end{table*}

To evaluate the performance of our framework for classifying Cyber threat tweets, we trained our classifier with all the CVE descriptions released in 2017. We tested our classifier on the tweets posted in 2018 from 50 influential Cyber security related accounts on Twitter. 
All of these Twitter accounts are known as experts or organizations working in the Cyber security domain and each of them has more than 5000 followers. Table \ref{tbl:accounts} lists the 50 Twitter accounts for collecting Cyber threat tweets.

Since the total number of the tweets posted in 2018 from the 50 Twitter accounts is very large (76205 tweets), it is not practical to manually label all these tweets for verifying the classification performance.
Therefore, we selected only a subset of the posted tweets to create the testing data set.
To cover all the posted tweets that were potentially relevant to Cyber threats, we weighted the relevance of each tweet to Cyber threats and selected only the tweets with high relevance score.

Because the training data consisted of only Cyber threat descriptions, we assumed that the more frequently a term appears in the training data set, the more relevant is the term to Cyber threats.
The more relevant to Cyber threats a term is, the larger the relevance weight of the term is. 
Therefore, we defined the relevance weight of a term t to Cyber threats as: 

\begin{equation}
\label{eq01}
rw(t) = log(1+\frac{n_t}{N-n_t+1})
\end{equation}
where 
$N$ is the total number of documents in the training data set
and $n_t$ is the total number of the documents  that contain term $t$ in the training data set.

\begin{figure}[t!]		
	\includegraphics[scale=0.42]{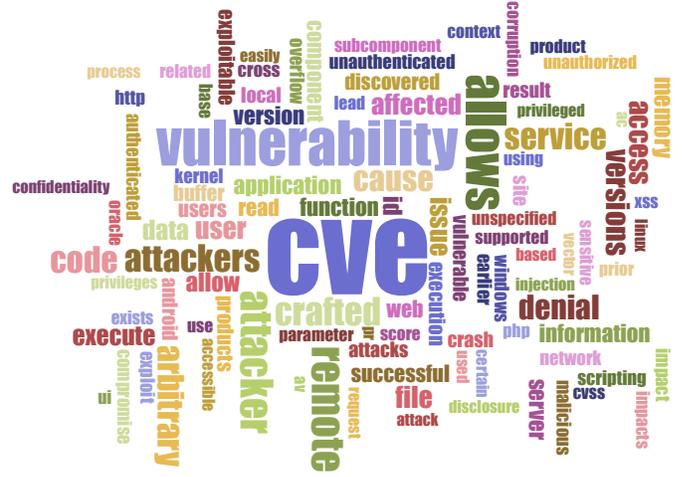}
	\caption{Word cloud of the top 100 popular terms in our training data set}
	\label{fig:top100}
\end{figure}

Fig.~\ref{fig:top100} illustrates the word cloud of the top 100 popular terms in the training data set. 
It can be inferred from the figure that the terms such as 'CVE' and 'vulnerability' have larger relevance weights to Cyber threats than the other terms such as 'service' and 'versions'.

The relevance weight of a tweet to Cyber threats can be calculated as the sum of the relevance scores of all the terms, weighted by their occurrences, in the tweet \cite{domeniconi2015comparison}. Therefore, we defined the relevance weight of a tweet d to Cyber threats as

\begin{equation}
\label{eq02}
RW(d) = \sum_{t \in d} f(t,d) * rw(t)
\end{equation}

where t is a term in tweet d, $f(t,d)$ is the number of the occurrences of term t in tweet d
and $rw(t)$ is the relevance weight of term t to Cyber threats.

Combining \eqref{eq01} and \eqref{eq02}, the relevance weight of a tweet d to Cyber threats can be rewritten as

\begin{equation}
RW(d) = \sum_{t \in d} f(t,d) * log(1+\frac{n_t}{N-n_t+1})
\end{equation}

We calculated the relevance weights for all the tweets posted in 2018 from the 50 Twitter users and selected only the top 3000 tweets with the highest relevance weight to create the testing data set. 
The selected tweets were then manually labeled as Cyber threat-relevant or irrelevant by two of the authors (one is a postdoctoral researcher and the other is a PhD student in the field relating to Cyber security). 
After labeling the selected tweets, we created a challenging testing data set with 232 tweets labeled as positive and 2768 tweets labeled as negative. 

\paragraph{Evaluation metrics}To measure the classification performance,
we used three common metrics including Precision, Recall and F1-score. The definitions of these metrics are given as follows. 
$$Precision = \frac{\textnormal{True positives}}{\textnormal{True positives + False positives}} $$
and 
$$Recall = \frac{\textnormal{True positives}}{\textnormal{True positives + False negatives}}$$
where True positives are the correctly classified Cyber threat-relevant tweets, False positives are the Cyber threat irrelevant tweets that are classified as relevant and False negatives are the Cyber threat-relevant tweets that are classified as irrelevant.

F1-score is a combination of Precision and Recall given by their harmonic mean. 
$$F1-score = \frac{\textnormal{2 * Precision * Recall}}{\textnormal{Precision + Recall}}$$

\subsection{Results and discussions}	
\begin{figure}[t!]
	\begin{center}		
		\includegraphics[]{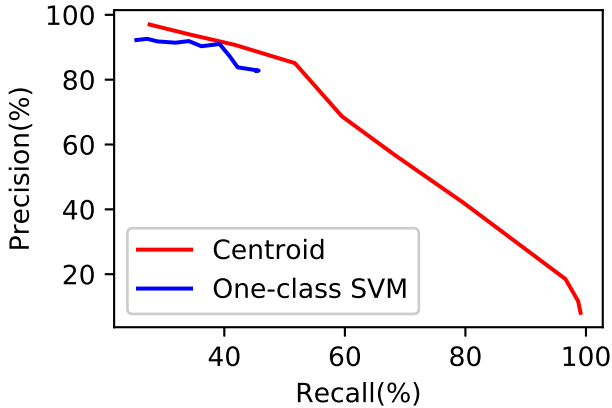}
		\caption{Precision as a function of Recall when varying the decision threshold of the Centroid and One-class SVM classifiers}
		\label{fig:recall_precision}
	\end{center}
\end{figure}
	
Fig.~\ref{fig:recall_precision} plots Precision as a function of Recall achieved by the Centroid and the One-class SVM classifiers. Precision and Recall are computed by varying the threshold parameter of these classifiers for deciding whether a tweet is normal or anomalous to Cyber threats. 
Normal tweets are labeled as Cyber threat-relevant while anomalous tweets are labeled as Cyber threat-irrelevant.
As can be seen from the figure, the Centroid classifier achieves a higher Precision rate than the One-class SVM classifier at the same Recall rate. 
This means that the Centroid classifier detects less number of false positives than the One-class SVM classifier providing that both the classifiers give the same number of true positives. 
The best overall performance, in term of F1-score, is 0.643 given by the Centroid classifier corresponding to the Precision value of 0.851 and the Recall value of 0.517. In further analysis, we used the Centroid classifier with the threshold parameter value that resulted in these Precision and Recall rates.

\paragraph{Comparison with baselines}
To show the effectiveness of our classification framework, we further compared our classifier with several baselines. 
The first baseline is the collection method of Cyber threat tweets based on the CVE identifier \cite{sabottke2015vulnerability}. 
This collection method simply collects only the tweets that contain the CVE identifier and ignores the tweets that do not have a CVE identifier.
Applying to our testing data set, 61 tweets with CVE identifier were collected but only 53 of them were relevant to Cyber threats.
Recalled that the total number of Cyber threat-relevant tweets in our testing data set was 232.
Therefore, collecting the Cyber threat tweets based on the CVE identifier gave the Precision rate of 53/61 ($\approx$0.869) and the Recall rate of 53/232 ($\approx$0.228). 
The F1-score given from these Precision and Recall values is 0.361, which is significantly below the F1-score of 0.643 achieved by our classifier.

We also compared our classifier with other baselines including Support Vector Machine (SVM), Multilayer Perceptron (MLP) and Convolutional Neural Network (CNN) \cite{alves2019processing,dionisio2019cyberthreat}.
These baselines are binary classification models which require to be trained with both positive and negative samples.
To obtain the negative samples, we randomly collected 3000 tweets that were irrelevant to Cyber threats from the 50 Twitter accounts (the tweets were verified by the two authors who labeled the testing data set).
The implementation of SVM and MLP are provided in the  scikit-learn Python package. The implementation of CNN is provided in the TensorFlow Python package 
\footnote{The TensorFlow Python package can be downloaded at https://www.tensorflow.org}. 
All the binary classification models were trained and executed with default parameter values.

\begin{table}[t!]
	\caption{Performance of our novelty classifier and the binary classifiers SVM, MLP and CNN}
	\label{tbl:compare}
	\begin{center}
		\begin{tabular}{|c|c|c|c|}
			\hline
			\textbf{Classifier} & \textbf{Precision} & \textbf{Recall} & \textbf{F1-score} \\ \hline
			SVM & 0.653 & 0.608  & 0.629\\ \hline
			MLP & 0.638 & 0.578  & 0.606 \\ \hline
			CNN & 0.474 & 0.625 & 0.539 \\ \hline
			Our novelty classifier & 0.851 & 0.517  & \textbf{0.643} \\ \hline
		\end{tabular}
	\end{center}
\end{table}
Table \ref{tbl:compare} compares the performance of our novelty classifier and the binary classifiers SVM, MLP and CNN. 
It can be seen from Table \ref{tbl:compare} that the binary classifiers give a higher Recall rate than our classifier but have a notably lower Precision rate. 
In term of overall performance, our classifier achieves a higher F1-score than SVM, MLP and CNN.

\subsection{Analysis of classified tweets}	
\begin{table*}
	\caption{Examples of the tweets without CVE identifier that refer to a threat described by at least one of the top 10 most similar CVE descriptions.}
	\label{tbl:threat_tweet_CVE}		
	\begin{center}
		\begin{tabular}{|p{6cm}|p{8cm}|c|}
			\hline			
			\centering\textbf{Tweet} & \centering\textbf{CVE description} & \textbf{CVE ID} \\ \hline
			\begin{small}
				Newly Disclosed Cross-Site Scripting (XSS) Vulnerability Resides in the Popular \# CKEditor Rich-Text Editor Library That Comes Pre-Integrated in Drupal Core. [Rated Moderately Critical]    Affected Versions\_x0014\_ CKEditor 4.5.11 and later versions (Drupal 8 \&  7) 
			\end{small}				
			& \begin{small}
				Cross-site scripting (XSS) vulnerability in the Enhanced Image (aka image2) plugin for CKEditor (in versions 4.5.10 through 4.9.1; fixed in 4.9.2), as used in Drupal 8 before 8.4.7 and 8.5.x before 8.5.2 and other products, allows remote attackers to inject arbitrary web script through a crafted IMG element.   
			\end{small}
			& \begin{small}
				CVE-2018-9861 
			\end{small}
			\\ \hline
			\begin{small}
				DHCP client application that allows systems to automatically receive network parameters like IP addresses contains \# security vulnerability that allows \# hackers to run arbitrary commands
			\end{small}					
			& \begin{small}
				DHCP packages in Red Hat Enterprise Linux 6 and 7, Fedora 28, and earlier are vulnerable to a command injection flaw in the NetworkManager integration script included in the DHCP client. A malicious DHCP server, or an attacker on the local network able to spoof DHCP responses, could use this flaw to execute arbitrary commands with root privileges on systems using NetworkManager and configured to obtain network configuration using the DHCP protocol. 
			\end{small}	
			& \begin{small}
				CVE-2018-1111 
			\end{small}	
			\\ \hline
		\end{tabular}	
	\end{center}
\end{table*}	

\begin{table*}
	\caption{Examples of the tweets without CVE identifier that refer to a threat not described by any of the top 10 most similar CVE descriptions}
	\label{tbl:threat_tweet_no}
	\begin{center}
		\begin{tabular}{|p{17.2cm}|}
			\hline
			\textbf{Tweet} \\ 
			\hline
			\begin{small}					
				Cb TAU recently detected a \# Squiblydoo attack attempting to leverage regsvr32.exe \& scrobj.dll to download and execute scriptlet code via an \# XML file. This attack also attempts to use taskeng.exe and the schedule service as persistence mechanisms via 
			\end{small}
			\\ \hline
			\begin{small}				
				The Sharpshooter technique can allow attackers to use a script to run a .NET binary directly from memory  not ever needing to reside on disk. Using durable  AMSI-aided detection  Windows Defender ATP disrupts campaigns and a steady hum of daily activity. 					
			\end{small}
			\\ \hline
		\end{tabular}
	\end{center}
\end{table*}	

To demonstrate the usefulness of our classification method, we examined the relationship between the correctly classified Cyber threat-relevant tweets and threat descriptions in the CVE database.
Our classifier correctly labeled 120 Cyber threat-relevant tweets out of the 232 Cyber threat-relevant tweets in the training data set.
Out of the 120 correctly labeled Cyber threat-relevant tweets, 39 tweets contained the CVE identifier and 81 tweets did not.
Since the recent research  has well analyzed Cyber threat-relevant tweets with CVE identifier \cite{sabottke2015vulnerability,alves2019processing,dionisio2019cyberthreat}, we focus our analysis on only Cyber threat-relevant tweets without CVE identifier.

For each of the 81 Cyber threat-relevant tweets without CVE identifier, we collected the top 10 CVE descriptions which were most similar to the tweet \footnote{The similarity between a tweet and a CVE description was calculated by the cosine similarity measure} \footnote{The tweets were compared with only the CVE descriptions publicly disclosed between 01/01/2015 and 30/04/2019}.
Our annotators were then asked to identify that if each of the Cyber threat-relevant tweets refers to the same threat with at least one of the top 10 CVE descriptions.
We find that 34 of the 81 Cyber threat-relevant tweets without CVE identifier refer to the same threat with at least a CVE description.
Table \ref{tbl:threat_tweet_CVE} lists some examples of these  tweets and the corresponding CVE description. 
The other 47 Cyber threat-relevant tweets without CVE identifier refer to a threat that is not described by the top 10 CVE descriptions.
Table \ref{tbl:threat_tweet_no} lists some examples of these tweets.

Our analysis of the classification results suggests that Cyber threat-relevant tweets on Twitter do not often include the CVE identifier of the related threats. 
However, the related CVE identifier of a Cyber threat-relevant tweet can be identified by matching the tweet with the top 10 most similar CVE descriptions.
The matched CVE description, therefore, provides additional information that are valuable for Cyber threat-related applications such as exploit prediction \cite{sabottke2015vulnerability} and Indicators of Compromise (IoCs) generation \cite{dionisio2019cyberthreat}.


\section{Conclusion}
In this paper, we proposed an automated framework for gathering Cyber threat intelligence from Twitter. 
Our collection framework utilizes a novelty detection model
that learns the features of Cyber threat intelligence from the CVE descriptions and classifies each input tweet as either normal or anomalous to the class of Cyber threat intelligence.
We evaluated our framework on a challenging data set of the tweets collected over the twelve months of 2018 from 50 influential Cyber security related accounts.  
Our classifier achieved a performance of 0.643 measured by F1-score for classifying Cyber threat-relevant tweets, which is higher than the performance of several baselines including SVM, MLP and CNN. 
Our analysis on the correctly classified Cyber threat-relevant tweets suggests that these tweets do not often mention the CVE identifier of the related threats. 
Collecting these tweets and finding the related CVE identifier, therefore, provide further information that are valuable for Cyber threat-related applications.

For the future work, our classification framework for Cyber threat-relevant tweets can  be potentially enhanced by combining it with word embeddings \cite{le2014distributed,pennington2014glove} for feature extraction. 
The classification performance can also be improved by adding a phase of Named Entity Recognition (NER) for vulnerability-related entities to the current framework.

\section*{Acknowledgment}	
The work has been supported by the Cyber Security Research Centre Limited whose activities are partially funded by the Australian Government’s Cooperative Research Centres Programme.


\end{document}